\documentclass[preprint,amsmath,amssymb,subeqn,showpacs]{revtex4}

\usepackage{graphicx}
\usepackage{dcolumn}
\usepackage{bm}

\newcommand{\be}{\begin{eqnarray}}
\newcommand{\en}{\end{eqnarray}}
\newcommand{\bes}{\begin{subequations}}
\newcommand{\ens}{\end{subequations}}
\newcommand{\ben}{\begin{eqnarray*}}
\newcommand{\enn}{\end{eqnarray*}}
\newcommand{\pa}{\partial}
\newcommand{\na}{\nabla}
\newcommand{\f}{\frac}

\newcommand{\bi}{\begin{itemize}}
\newcommand{\ei}{\end{itemize}}

\newcommand{\ti}{\times}
\newcommand{\ot}{\otimes}
\renewcommand{\a}{\alpha}
\renewcommand{\b}{\beta}
\renewcommand{\r}{\rho}
\newcommand{\As}{\alpha_{\lambda}}
\newcommand{\Cs}{{\chi}_{\lambda}}
\newcommand{\Ar}{\alpha_{\rho}}
\newcommand{\Cr}{{\chi}_{\rho}}
\newcommand{\s}{\lambda}
\renewcommand{\c}{\chi}
\newcommand{\Qs}{{\bf{Q}}_{\lambda}}
\newcommand{\Qr}{{\bf{Q}}_{\rho}}
\newcommand{\Qx}{{\bf{Q}}_{\xi}}
\newcommand{\Qc}{{\bf{Q}}_{c}}
\newcommand{\Qp}{{\bf{Q}}_{\psi}}
\newcommand{\R}{\Rightarrow}

\begin{document}
\title{On dynamics of velocity vector potential in incompressible fluids}
\author{Sagar Chakraborty}
\email{sagar@bose.res.in}
\affiliation{S.N. Bose National Centre for Basic Sciences, Saltlake, Kolkata 700098, India}
\author{Partha Guha}
\email{partha@bose.res.in}
\affiliation{S.N. Bose National Centre for Basic Sciences, Saltlake, Kolkata 700098, India}
\affiliation{Max Planck Institute for Mathematics in the Sciences, Inselstrasse 22, D-04103 Leipzig, Germany}
\date{\today}
\begin{abstract}
An elegant quaternionic formulation is given for
the Lagrangian advection equation for velocity vector potential in fluid dynamics. 
At first we study the topological significance of
a restricted conserved quantity {\it viz.,} stream-helicity and later more realistic configuration 
of open streamlines is figured out.
Also, using Clebsch parameterisation of the velocity vector potential 
yet another physical significance for the stream-helicity is provided. Finally we give
a Nambu-Poisson formalism of the Lagrangian advection equation for velocity vector potential.
\end{abstract}
\maketitle
\section{INTRODUCTION}
Quite recently\cite{Sagar}, a conserved quantity--- stream-helicity--- has been conjectured for certain cases of incompressible fluid flows.
Although Arnold and Khesin \cite{Arnold} have documented in their book a rather general topological study of quantities of such type, stream-helicity has never been explicitly studied and used in the mathematical fluid dynamics.
Its study becomes even more important at the face of the discovery that it has been shown to remain conserved for at least some kinds of fluid flows.
Therefore, an upsurge of motivation to study stream-helicity and the dynamics and beautiful mathematical aspects of the velocity vector potential (that has got itself involved in defining the stream-helicity) is quite natural.
\\
In this paper, as a rather interesting product of the study, the possible quaternionic structure of the Lagrangian advection equation for the velocity vector potential has been unleashed; readers may be aware that quaternions, a concept originally invented by Hamilton to generalise complex numbers to $\mathbb{R}^4$, obey multiplication rules governed by a simple non-commutative division algebra and currently play an important part in the theory of 4-manifolds, the technical foundation of modern inertial guidance systems in the aerospace industry, the control of the orientation of tumbling objects in computer animations {\it etc}.
Also, for the aforementioned advection equation we have given a possible Nambu-Poisson formulation.
Most importantly, the topological meaning for the stream-helicity has been extended to unclosed but tangled streamlines.
For the sake of clarity and completeness, we have briefly but appropriately summarised included the essence of the paper (\cite{Sagar}) herein in the section (IV).
\\
This paper is {\bf organised} as follows: Section II deals with the advection equation 
of the velocity vector potential. We give the quaternionic formulation of the advection equation 
in Section III.  
Section IV is devoted on stream-helicity starting right from its definition and Section V 
discusses the stream-helicity's geometrical and topological interpretations for open streamlines.
We indroduces Clebsch parameterisation for the vector potential in Section VI.  
Section VII reformulates the advection equation for the velocity vector potential using Nambu brackets. 
\section{ADVECTION EQUATION FOR VELOCITY VECTOR POTENTIAL}
Let us begin with the Euler equation
\be
\f{\pa{\vec{u}}}{\pa t}+({\vec{u}}.{\vec{\na}}){\vec{u}}=-{\vec{\na}}P
\label{1}
\en
for three-dimensional, inviscid, incompressible fluid being acted upon by irrotational body forces.
The pressure $P$ used in the equation (\ref{1}) includes the effect of such forces also.
As we have assumed the fluid to be incompressible, {\it i.e.}, the density is constant, we have taken the density to be unity without the loss of generality.
The mathematical way of acknowledging incompressibility yields for the velocity field $\vec{u}$:
\be
{\vec{\na}}.{\vec{u}}=0
\label{2}
\en
paving way for defining the vector potential $\vec{\xi}$ for the velocity field as follows:
\be
{\vec{u}}={\vec{\na}}\times{\vec{\xi}}
\label{3}
\en
Obviously, $\vec{\xi}$ is not unique, for, a term $\vec{\na}\gamma$, $\gamma$ being a scalar 
field can always be added to it keeping $\vec{u}$ unchanged.
We shall come back to this issue in the end of Section IV.
For the time being we put relation (\ref{3}) in the equation (\ref{1}) to get:
\be
\f{\pa}{\pa t}({\vec{\na}}\times{\vec{\xi}})+({\vec{u}}.{\vec{\na}})
({\vec{\na}}\times{\vec{\xi}})=-{\vec{\na}}P
\label{4}
\en
Trivial manipulations using vector algebra suggests:
\be
[{\vec{\na}}\times({\vec{u}}.{\vec{\na}}){\vec{\xi}}]_i&=&\epsilon_{ijk}\pa_j(u_l\pa_l\xi_k)\nonumber\\
\R[{\vec{\na}}\times({\vec{u}}.{\vec{\na}}){\vec{\xi}}]_i&=&\epsilon_{ijk}(\pa_ju_l)(\pa_l\xi_k)+
\epsilon_{ijk}u_l\pa_j\pa_l\xi_k\nonumber\\
\R[({\vec{u}}.{\vec{\na}})({\vec{\na}}\times{\vec{\xi}})]_i&=&[{\vec{\na}}\times({\vec{u}}.{\vec{\na}}){\vec{\xi}}]_i-\epsilon_{ijk}(\pa_ju_l)
(\pa_l\xi_k)
\label{5}
\en
Putting relation (\ref{5}) in the equation (\ref{4}) we get the following Lagrangian advection equation for the velocity vector potential:
\be
\f{\pa{\vec{\xi}}}{\pa t}+({\vec{u}}.{\vec{\na}}){\vec{\xi}}=\textrm{curl}^{-1}{\vec{\eta}}
\label{6}
\en
where $\vec{\eta}$ is defined as:
\be
{\eta}_i\equiv \epsilon_{ijk}T_{jk} - \pa_iP, 
\label{7}
\en
where we interpret $T_{jk}$ as a contribution of a stress tensor which, of course, is not symmetric.
The stress tensor $T_{jk}$ that does all the possible deformations {\it viz.}, tilt, stretch, rotation {\it etc.} of the velocity vector potential lines has been defined as
\be
T_{jk}\equiv(\pa_ju_l)(\pa_l\xi_k)
\label{7a}
\en
Definitely $\textrm{curl}^{-1}{\vec{\eta}}$ is unique only upto a gradient of a scalar, say, 
${\theta}$, an issue taken up again in the Section IV.
So, we have yet another gauge to be fixed.
In the next section, we shall note the equation (\ref{6}), is endowed with a beautiful exposition of
quaternionic structure {\it $\acute{\textrm{a}}$ la} Gibbon {\it et. al.}\cite{Gibbon1,Gibbon2,Gibbon3}\\
\section{QUATERNIONIC FORMULATION FOR ADVECTION EQUATION OF $\vec{\xi}$}
Let us define $\vec{\s}\equiv\textrm{curl}^{-1}\vec{\eta}$ for the sake of simplicity of writing.
Of course, as discussed in the Section II, we need to fix a gauge for $\vec{\s}$ and this possibly can be done as in the many standard ways of gauge fixing. 
Thus, we can rewrite the equation (\ref{6}) and its Lagrangian time derivative respectively as
\begin{subequations}
\be
\f{D\vec{\xi}}{Dt}=\vec{\s}
\label{3.1}
\en
\be
\f{D^2\vec{\xi}}{Dt^2}=\vec{\r}\equiv\f{D\vec{\s}}{Dt}
\label{3.2}
\en
\end{subequations}
where $\vec{\r}$ has been appropriately defined.
Unlike the vorticity vector in the ideal incompressible fluid, velocity vector 
potential is not known to be frozen in the fluid.
However, it doesn't prevent lines of $\vec{\xi}$ from getting tilted, stretched 
and rotated due to the fluid motion; rather the roller-coaster ride of each 
line of $\vec{\xi}$ is even more complex than the usually studied field lines 
of the vorticity vector;  $\vec{\omega}\equiv\vec{\na}\ti\vec{u}$.
Evidently, $\vec{\s}$ is the relevant vector field which brings about the 
deformation in the $\vec{\xi}$ as the fluid evolves.
The information about how $\vec{\xi}$ is deformed is obviously kept hided in the 
evolution of the relative angle and magnitude of $\vec{\xi}$ with respect to $\vec{\s}$.
Keeping this in mind we define following quantities:
\be
\a_{\s}\equiv\f{\vec{\xi}.\vec{\s}}{\vec{\xi}.\vec{\xi}},\phantom{x}\vec{\c}_{\s}\equiv\f{\vec{\xi}\times\vec{\s}}{\vec{\xi}.\vec{\xi}},\phantom{x}\a_{\r}\equiv\f{\vec{\xi}.\vec{\r}}{\vec{\xi}.\vec{\xi}},\phantom{x}\vec{\c}_{\r}\equiv\f{\vec{\xi}\times\vec{\r}}{\vec{\xi}.\vec{\xi}}
\label{3.3}
\en
Also, if the instantaneous angle between $\vec{\xi}$ and $\vec{\s}$ be $\phi_{\s}$, then 
\be
&&\tan{\phi_{\s}}\equiv\f{|\vec{\xi}\times\vec{\s}|}{\vec{\xi}.\vec{\s}}=\f{\c_{\s}}{\a_{\s}}\label{3.4}\\
\R&&\s^2=\xi^2\a^2_{\s}\sec^2{\phi_{\s}}\label{3.5}
\en
Using the relations (\ref{3.3}) and (\ref{3.5}), following equations may be constructed:
\begin{subequations}
\be
\f{D}{Dt}(\vec{\xi}.\vec{\s})=\xi^2(\As^2\sec^2\phi_{\s}+\Ar)
\label{3.6a}
\en
\be
\f{D}{Dt}(\vec{\xi}\ti\vec{\s})=\xi^2\vec{\chi}_{\r}
\label{3.6b}
\en
\end{subequations}
which can then be modified respectively into the following Lagrangian advection relations for $\As$ and $\vec{\chi}_{\s}$:
\begin{subequations}
\be
\f{D\As}{Dt}=\Cs^2-\As^2+\Ar
\label{3.7a}
\en
\be
\f{D\vec{\chi}_{\s}}{Dt}=-2\As\vec{\chi}_{\s}+\Cr
\label{3.7b}
\en
\end{subequations}
The equation (\ref{3.7b}) may be rewritten for the magnitude of $\vec{\chi}_{\s}$ as:
\be
\f{D\Cs}{Dt}=-2\As\Cs+\Cr'
\label{3.8}
\en
\be
\textrm{where,}\phantom{xxx}\Cr'\equiv\f{\vec{\chi}_{\s}.\vec{\chi}_{\r}}{\Cs}
\label{3.9}
\en
Again, the equations (\ref{3.1}) and (\ref{3.2}) when recasted, with the help of the definitions (\ref{3.3}) and their evolution equations (\ref{3.7a}) and (\ref{3.7b}), to get Lagrangian dynamics for the magnitude of $\vec{\xi}$, one arrives at:
\begin{subequations}
\be
\f{D\xi}{Dt}=\As\xi
\label{3.10a}
\en
\be
\f{D^2\xi}{Dt^2}=(\Cs^2+\Ar)\xi
\label{3.10b}
\en
\end{subequations}
This highlights how significantly the dynamics of velocity vector potential is governed by the quantities defined in the relation (\ref{3.3}).\\ 
Now, let $\hat{\xi}$, $\hat{\chi}_{\s}$ and $\hat{\xi}\ti\hat{\chi}_{\s}$ be the unit vector along $\vec{\xi}$, $\vec{\chi}_{\s}$ and $\vec{\xi}\ti\vec{\chi}_{\s}$ respectively.
One can form an orthonormal frame $(\hat{\xi},\hat{\xi}\ti\hat{\chi}_{\s},\hat{\chi}_{\s})$ using them.
We wish to study the Lagrangian dynamics of the frame.
In this frame one can apparently write the vector $\vec{\r}$ in terms of the independent unit vectors $\hat{\xi}$, $\hat{\xi}\ti\hat{\chi}_{\s}$ and $\hat{\chi}_{\s}$ as
\be
&&\vec{\r}=\Ar\hat{\xi}+a(\hat{\xi}\ti\hat{\chi}_{\s})+b\hat{\chi}_{\s}
\label{3.11}\\
\R&&\vec{\chi}_{\r}=b(\hat{\xi}\ti\hat{\chi}_{\s})-a\hat{\chi}_{\s}
\label{3.12}
\en
We have cross-producted the equation (\ref{3.11}) with $\hat{\xi}$ to reach at the relation (\ref{3.12}).
Therefore, obviously $a=-(\hat{\chi}_{\s}.\vec{\chi}_{\r})$ and $b=\hat{\xi}.(\hat{\chi}_{\s}\ti\vec{\chi}_{\r})$.
Now, defining the Darboux angular velocity vector as
\be
\vec{\Omega}_D\equiv\vec{\chi}_{\s}+\f{b}{\Cs}\hat{\xi}
\label{3.13}
\en
we can arrive (using the expressions (\ref{3.1}), (\ref{3.3}), (\ref{3.7b}) and (\ref{3.8})) at the following evolution equations determining the Lagrangian dynamics of the frame:
\bes
\be
\f{D\hat{\xi}}{Dt}=\vec{\Omega}_D\ti\hat{\xi}
\label{3.14a}
\en
\be
\f{D}{Dt}\left(\hat{\xi}\ti\hat{\chi}_{\s}\right)=\vec{\Omega}_D\ti\left(\hat{\xi}\ti\hat{\chi}_{\s}\right)
\label{3.14b}
\en
\be
\f{D\hat{\chi}_{\s}}{Dt}=\vec{\Omega}_D\ti\hat{\chi}_{\s}
\label{3.14c}
\en
\ens
Before delving into the quaternionic formulation let us see how complex numbers, precursor of quaternions, can enrich the beauty of the equations being dealt with.
Lets define
\be
\zeta_{\s}=\As+i\Cs,\phantom{xxx}\zeta_{\r}=\Ar+i\Cr
\label{3.15b}
\en
After a little bit of manipulation, putting the expressions (\ref{3.7a}) and (\ref{3.8}) into use, we can arrive at
\be
\f{D\zeta_{\s}}{Dt}+\zeta_{\s}^2-\zeta_{\r}=0
\label{3.16}
\en
which can be decomposed into the following two equations
\bes
\be
\f{D|\zeta_{\s}|^2}{Dt}+2\As|\zeta_{\s}|^2-2\Re[\zeta_{\s}\bar{\zeta}_{\r}]=0
\label{3.17a}
\en
\be
\f{D}{Dt}(\tan\phi_{\s})+\As\tan^3\phi_{\s}+\left(\As+\f{\Ar}{\As}\right)\tan\phi_{\s}-\f{\Cr'}{\As}=0
\label{3.17b}
\en
\ens
where the `overhead bar' denotes the operation of complex conjugation.
By the way, by picking a function $\psi$, such that $\zeta_{\s}=\f{1}{\psi}\f{D\psi}{Dt}$, one can linearise the equation (\ref{3.16}) to the form:
\be
\f{D^2\psi}{Dt^2}-\zeta_{\r}\psi=0
\label{3.18}
\en
Mood is now set up for introducing the quaternionic structure of the equations for the variables $\As,\Cs,\Ar,\Cr$ and $\xi$.
We begin by defining following quaternions as the natural tetrads
\be
\Qs\equiv[\As,\vec{\chi}_{\s}],\phantom{x}\Qr\equiv[\Ar,\vec{\chi}_{\r}]\phantom{x}\textrm{and}\phantom{x}\Qx\equiv[0,\vec{\xi}]
\label{3.19}
\en
In what follows let `$\ot$' and `asterisk' denote the Grassman product and the quaternion conjugate respectively.
Then obviously,
\be
\Qs\ot\Qx=[\As,\vec{\chi}_{\s}]\ot[0,\vec{\xi}]=[-\vec{\chi}_{\s}.\vec{\xi},\As\vec{\xi}+\vec{\chi}_{\s}\ti\vec{\xi}]
\label{3.20}
\en
So, one has by mere inspection
\be
\f{D\Qx}{Dt}-\Qs\ot\Qx=0
\label{3.21}
\en
and similarly,
\be
\f{D^2\Qx}{Dt^2}-\Qr\ot\Qx=0
\label{3.22}
\en
Differentiating equation (\ref{3.21}) and putting equation (\ref{3.22}) into use in the midway of the elementary calculations, we get the compatibility condition:
\be
\f{D\Qs}{Dt}\ot\Qx+\Qs\ot\Qs\ot\Qx-\Qr\ot\Qx=0
\label{3.23}
\en
which due to associativity yields on taking Grassman product with $\Qx^{-1}$ from the right
\be
\f{D\Qs}{Dt}+\Qs\ot\Qs-\Qr=0
\label{3.24}
\en
that is nothing but the quaternionic Riccati equation that can be linearised using a quaternion $\Qp$ such that
\be
\Qs=\f{D\Qp}{Dt}\ot\Qp^{-1}
\label{3.25}
\en
where, $\Qp^{-1}=\f{\Qp^{\ast}}{\Qp\ot\Qp^{\ast}}$ is the inverse of $\Qp$.
Using this in the equation (\ref{3.23}) and the following identity for any arbitrary quaternion ${\bf Q}(\vec{x},t)$:
\be
\f{D{\bf Q}^{-1}}{Dt}=-{\bf Q}^{-1}\ot\f{D{\bf Q}}{Dt}\ot{\bf Q}^{-1}
\label{3.26}
\en
we arrive at the linearised quaternionic Riccati equation:
\be
\f{D^2\Qp}{Dt^2}-\Qr\ot\Qp=0
\label{3.27}
\en
Now, let us set up a Lagrangian advection equation for $\Qr$ analogous to the relation (\ref{3.24}).
To do that let us note that as both $\vec{\chi}_{\s}$ and $\vec{\chi}_{\r}$ are perpendicular to $\hat{\xi}$, one may write
\be
&&\hat{\xi}.\vec{\chi}_{\r}=0\label{qq}\\
\R&&\f{D\vec{\chi}_{\r}}{Dt}=\vec{\chi}_{\s}\ti\vec{\chi}_{\r}+c_1\vec{\chi}_{\s}+c_2\vec{\chi}_{\r}
\label{3.28}
\en
where $c_1$ and $c_2$ are two undetermined scalars.
In reaching the expression (\ref{3.28}) from the relation (\ref{qq}), we have made use of the equation (\ref{3.14a}).
Using the definition (\ref{3.3}) explicitly in the relation (\ref{3.28}), a bit of manipulation yields
\be
&&\hat{\xi}\ti\left[\f{D}{Dt}\left(\f{\vec{\r}}{\xi}\right)-\Ar\f{\vec{\s}}{\xi}\right]=c_1\vec{\chi}_{\s}+c_2\vec{\chi}_{\r}\\
\R&&\f{D}{Dt}\left(\f{\vec{\r}}{\xi}\right)=\Ar\f{\vec{\s}}{\xi}+c_1\vec{\chi}_{\s}\ti\hat{\xi}+c_2\vec{\chi}_{\r}\ti\hat{\xi}+c_3\hat{\xi}
\label{3.29}
\en
where $c_3$ is yet another unknown scalar.
Recalling the relations (\ref{3.3}) and (\ref{3.14a}), the preceding equation (\ref{3.29}) when dot-producted with $\hat{\xi}$ yields
\be
\f{D\Ar}{Dt}=\As\Ar+\vec{\chi}_{\s}.\vec{\chi}_{\r}+c_3
\label{3.30}
\en
With the information contained in the relations (\ref{3.28}) and (\ref{3.30}) one can get the coveted quaternion equation as follows:
\be
\f{D\Qr}{Dt}=\Qs\ot\Qr+\Qc
\label{3.31}
\en
In which, using the identity quaternion (${\bf I}=[1,0,0,0]$) with respect to the Grassman product, we have defined the quaternion $\Qc$ as
\be
\Qc\equiv c_1'\Qr+c_2'\Qs+c_3'{\bf I}
\label{3.32}
\en
$c_1'$, $c_2'$ and $c_3'$, in turn, are the redefined unknown scalar given as
\be
c_1'\equiv c_1-\As,\phantom{x}c_2'\equiv c_2-\Ar,\phantom{x}\textrm{and}\phantom{x}c_3'{\bf I}\equiv(c_3-c_1\Ar-c_2\As){\bf I}+\Qs\ot\Qr^{\ast}+\Qr\ot\Qs^{\ast}
\label{3.33}
\en
Thus, unfortunately we could find the Lagrangian quaternionic equation for $\Qr$ only at the cost of introducing three undetermined scalars.
\section{STREAM-HELICITY}
The `stream-helicity' ($S$) is defined as:
\be
S\equiv\int_V{\vec{\xi}}.{\vec{u}}d^3x
\label{8}
\en
where $V$ is a volume occupied by the fluid.
At this point let us ponder over the aforementioned non-uniqueness of the vector potential\cite{Berger}.
For smooth discussion's sake, we assume for the time being that the volume is simply connected.
Suppose $\vec{\xi}\rightarrow\vec{xi}+\vec{\na}\gamma$, then from the definition (\ref{8}) of 
stream-helicity we can find the change $\delta{S}$ in $S$ to be:
\be
\delta S=\int_V\vec{\na}\gamma.\vec{u}d^3x=\oint_{\pa{V}}\gamma\vec{u}.\hat{n}d^2x
\label{ds}
\en
where $\hat{n}$ is the unit vector perpendicular to the infinitesimal surface element $d^2x$ 
and we have used the relation (\ref{2}) and Gauss divergence theorem.
The relation (\ref{ds}) amounts to saying that the stream-helicity will be gauge invariant 
in case the surface $\pa V$ bounding $V$, is the surface made up of streamlines {\it i.e.}, 
$\vec{u}.\hat{n}=0$ on $\pa V$.
This condition for gauge invariance is rather strong because if $\vec{u}.\hat{n}\ne 0$ on 
$\pa V$ then one cannot seek refuge in Coulomb gauge for it is too loosely defined inside $V$ 
with no information about the outside field whatsoever.
More starkly, it means that different solenoidal vector potentials inside $V$ can correspond to 
Coulomb potentials of fields which have different structures outside $V$.
Now if we relax the condition that $V$ is simply connected, then the line integrals of 
$\vec{\xi}$ about the `holes' in the possibly multiply connected region have to be 
specified in order to have gauge-invariant stream-helicity within $\pa V$ on which $\vec{u}.\hat{n}=0$.
Another good thing about this very gauge is that it helps in bypassing the issue of 
non-uniqueness of $\textrm{curl}^{-1}{\vec{\eta}}$ in the discussion on the conservation 
of $S$ in this section by causing the R.H.S. of the expression (\ref{999}) to become 
independent of $\theta$ that naturally infects the second term of the R.H.S. of the equation.
\\
Let us take total derivative of $S$ w.r.t. time to get:
\be
\f{dS}{dt}&=&\int\f{D}{Dt}({\vec{\xi}}.{\vec{u}})d^3x\\
\Rightarrow\f{dS}{dt}&=&\int{\vec{\xi}}.(-{\vec{\na}}P)d^3x+\int{\vec{u}}.(\textrm{curl}^{-1}{\vec{\eta}})d^3x
\label{999}
\en
where, $D/Dt$ is the material derivative w.r.t. time and it basically is a shorthand for $\pa/\pa t+\vec{u}.\vec{\na}$.
Again, simple vector algebra suggests:
\be
({\vec{\na}}\times{\vec{\xi}}).(\textrm{curl}^{-1}{\vec{\eta}})={\vec{\na}}.({\vec{\xi}}\times\textrm{curl}^{-1}{\vec{\eta}})+{\vec{\eta}}.{\vec{\xi}}
\label{101010}
\en
With relation (\ref{3}) in mind, inserting relation (\ref{101010}) in the equation (\ref{999}), we have the following:
\be
\f{dS}{dt}&=&-2\int{\vec{\xi}}.{\vec{\na}}Pd^3x+\int{\vec{\na}}.({\vec{\xi}}\times\textrm{curl}^{-1}{\vec{\eta}})d^3x+\int\xi_i\pa_l(\epsilon_{ijk}\xi_k\pa_ju_l)d^3x
\label{111111}
\en
where, equation (\ref{2}) has been used.
If all the terms in the R.H.S. of equation (\ref{111111}) vanish then one may set
\be
\f{dS}{dt}=0
\label{12}
\en
and say that stream-helicity is a conserved quantity.
\\
\cite{Sagar} picks it up right from here to conjecture that in certain kind of fluid flow the stream-helicity is conserved.
Further, it has been shown therein that not only does this remain conserved but also one can lend stream-helicity a topological interpretation for the closed linked stream-tubes and knotted stream-tubes; the stream-helicity is proportional to the linkage number in the case of linked stream-tubes for Beltrami vector potential and in the case of knotted tubes it is connected to the way in which the knot may be constructed from a closed tubular unknotted stream-tube.
\\
Historically speaking, after almost hundred years of Lord Kelvin's realisation that any linkage or any knottedness in the vorticity field at any earlier time should remain conserved at all later times in an inviscid and barotropic fluid being acted upon by irrotational body forces, Moreau\cite{1} and later Moffatt\cite{2} established an invariant known as helicity which is of topological character and encompasses Kelvin's insight.
But what Lord Kelvin had missed was the possible existence of knotted streamtubes in the steady Euler flows --- a fact very logically speculated by Moffatt\cite{3}.
Actually, inspired by this speculation only, a new quantity {\it viz.}, stream-helicity has been introduced for the case of inviscid and incompressible fluid being forced by irrotational body forces.
\section{TOPOLOGICAL MEANING OF $S$ FOR UNCLOSED STREAMLINES}
In this paper we show that this analogy may be stretched further to good use.
Just as Arnold\cite{Hopf} extended the interpretation the linking of the oriented vortex tubes introduced by Moffatt to the situations in which vortex lines are not closed curves but are wound about each other infinitely often, we too can identify the stream-helicity as the ``asymptotic Hopf invariant''.
Probably the physical situations of the tornado, whirlpool etc. spiraling flows that may cover quite a distance in space can entertain such an interpretation for the stream-helicity, for, the existence of the streamlines wound many times around each other is a rather strong possibility therein.\\
Let us try to be more explicit in the language of differential geometry.
For the case of incompressible flows being dealt with herein, let the velocity field $\vec{u}$ be the field on a 3-D closed oriented simply connected manifold $\mathbb{M}$ with a volume form $\mu$ and the phase flow of $\vec{u}$ be $\phi^t:\mathbb{M}\rightarrow\mathbb{M}$.
Let $\mathbb{M}$ has the real cohomology of a three-sphere which means $H^1(\mathbb{M})=H^2(\mathbb{M})=0$.
Then, one knows\cite{Vogel} a set $\Sigma$ containing exactly one oriented path having starting point $p$ and end point $q$ for every pair $(p,q)$ of points of $\mathbb{M}$ is always possible to choose.
As we are concerned with the realistic fluids, in what follows it will suffice for us to consider $\mathbb{M}$ to be 3-D compact Euclidean domain ($\mathbb{M}\subset\mathbb{R}^3$) such that $\vec{u}$ is tangent to the boundary $\pa\mathbb{M}$ which is, by the way, required for the uniqueness of stream-helicity as discussed at the end of the section (II).
With these in mind, we consider a pair of points $x,y\in\mathbb{M}$ and select two large numbers $T_x$ and $T_y$ so that the trajectories $\phi^t(x)\phantom{x}(t\in[0,T_x])$ and $\phi^t(y)\phantom{y}(t\in[0,T_y])$ are closed by joining $x$ and $\phi^{T_x}(x)$ (and similarly, $y$ and $\phi^{T_y}(y)$) using a path belonging to $\Sigma$.
Let the two closed curves formed be $\Gamma_x$ and $\Gamma_y$ respectively.
It is assumed that $\Gamma_x$ and $\Gamma_y$ do not interest.
Obviously, $\Gamma_x:S^1\rightarrow\mathbb{M}$ and $\Gamma_y:S^1\rightarrow\mathbb{M}$ are smooth mappings of two circumferences to $\mathbb{M}$ with disjoint images.
In the images, $\dot{\Gamma}_x(t_x)$ and $\dot{\Gamma}_y(t_y)$ are the velocity vectors found by differentiating the curves with respect to the corresponding parameters $t_x$ ($mod\phantom{x}T_x$) and $t_y$ ($mod\phantom{x}T_y$) respectively.
One can now define linking number for the closed curves $\Gamma_x(S^1)$ and $\Gamma_y(S^1)$ in $\mathbb{M}$ by the Gauss theorem (see {\it e.g.}, \cite{Arnold}) as:
\be
lk(\Gamma_x,\Gamma_y)=\f{1}{4\pi}\int_0^{T_x}\int_0^{T_y}\f{(\dot{\Gamma}_x(t_x),\dot{\Gamma}_y(t_y),\Gamma_x(t_x)-\Gamma_y(t_y))}{||\Gamma_x(t_x)-\Gamma_y(t_y)||^3}dt_xdt_y
\label{Gauss}
\en
where $(\cdot,\cdot,\cdot)$ and $||\cdot||$ are the symbols for vector triple product and the norm respectively.
Now, the asymptotic linking number for $\phi^t(x)$ and $\phi^t(y)$ of $\vec{u}$ is traditionally defined as:
\be
\lambda_{\vec{u}}(x,y)=\lim_{T_x,T_y\rightarrow\infty}\f{lk(\Gamma_x,\Gamma_y)}{T_xT_y}
\label{aln}
\en
The limit is known to be existing almost everywhere.
This relation (\ref{aln}) basically the Gauss linking number of $\phi^t(x)$ and $\phi^t(y)$ {\it i.e.},
\be
\lambda_{\vec{u}}(x,y)=\left(\f{1}{4\pi}\right)\lim_{T_x,T_y\rightarrow\infty}\int_0^{T_x}\int_0^{T_y}\f{(\dot{x}(t_x),\dot{y}(t_y),x(t_x)-y(t_y))}{||x(t_x)-y(t_y)||^3}dt_xdt_y
\label{gln}
\en
where $x(t)=\phi^t(x)$ and $y(t)=\phi^t(y)$ for convenience.
Subsequently, the average self-linking number of $\vec{u}$ may be proposed as:
\be
\lambda_{\vec{u}}=\int_{\mathbb{M}}\int_{\mathbb{M}}\lambda_{\vec{u}}(x,y)\mu_1\mu_2
\label{asln}
\en
One may now invoke helicity theorem due to Arnold\cite{Hopf} for the case of solenoidal velocity field $\vec{u}$ to seek a topological interpretation for the stream-helicity for the fluid flow having unclosed streamlines as:
\be
S=\lambda_{\vec{u}}
\label{finally}
\en
This completes our discussion on the possible topological interpretation for the restrictedly conserved quantity stream-helicity for the certain classes of flow.
\section{CLEBSCH PARAMETRISATION OF $\vec{\xi}$}
\indent In the form notation, $\vec{\xi}$ is a 1-form such that the Clebsch parameterisation of the velocity potential $\vec{\xi}$ using three independent scalars $\a$, $\b$ and $\gamma$ can be represented as:
\be
\xi=\a{d}\b+d\gamma
\label{xi1}
\en
The parameterisation is attainable locally in space owing to the Darboux's theorem.
Of course, the uniqueness of $\a$ and $\b$ is at stake because the expression (\ref{xi1}) keeps its form intact after any canonical transformation of $\a$ and $\b$.
In the standard vectorial notation, we have from (\ref{xi1}):
\be
\vec{\xi}&=&\a\vec{\na}\b+\vec{\na}\gamma
\label{xi2}\\
\R\vec{u}&=& \vec{\na}\times\vec{\xi}=\vec{\na}\a\times\vec{\na}\b
\label{v1}
\en
The expression (\ref{v1}) regales itself with the standard interpretation that each streamline is the line of intersection of a pair of surfaces (rather stream-surfaces) given by $\a=\textrm{constant}$ and $\b=\textrm{constant}$.
Now, the relations (\ref{xi2}) and (\ref{v1}) can be dotted to get the following form for the `stream-helicity density' $s$:
\be
s\equiv\vec{\xi}.\vec{u}=\vec{\na}.(\gamma{\vec{u}})
\label{sigma}
\en
This expression suggests that for the non-vanishing stream-helicity and appropriately well-behaved velocity field, $\gamma$ must be singular either on the surface at infinity bounding the integration region or in the finite volume of the integration domain.
Therefore, the stream-helicity becomes an integral over the surface bounding the singularities:
\be
\int_V\vec{\xi}.\vec{u}d^3x=\int_{\pa{V}}\{(\gamma{\vec{u}}).\hat{n}\}d^2x
\label{sing}
\en
As a result, a physical interpretation of the stream-helicity is that it is the measure of the flux of the velocity field, weighted by $\gamma$, through the surfaces that enclose the singularities of $\gamma$.
\\
\indent Again from the relations (\ref{v1}) and (\ref{sigma}), we can see that
\be
(\vec{u}.\vec{\na})\a=0;\phantom{x}(\vec{u}.\vec{\na})\b=0;\phantom{x}(\vec{u}.\vec{\na})\gamma=s
\label{threerel}
\en
which dictates that the Jacobian $J$ of the transformation (assumed invertible)
\be
\vec{x}=\vec{X}(\a,\b,\gamma,t)
\label{X}
\en
is given by:
\be
J\equiv\f{\pa(x,y,z)}{\pa(\a,\b,\gamma)}=\f{1}{\left\{(\vec{\na}\a\times\vec{\na}\b).\vec{\na}\gamma\right\}}=\f{1}{s}
\label{J1}
\en
Thus, the stream-helicity density $s$ is the Jacobian for the inverse transformation of $\vec{X}$ in (\ref{X}).
\section{NAMBU STRUCTURE AND HAMILTONIAN FORMULATION}
In 1973, in a remarkable paper, Nambu \cite{na} introduced a generalization of classical Hamiltonian mechanics which is now called Nambu mechanics. In his formalism, he replaced the canonical pair of variables found in Hamiltonian mechanics with a triplet of coordinates in an odd dimensional space. He formulated his dynamics in terms of ternary operation, the Nambu bracket, as opposed to the usual binary Poisson bracket. The original motivation of Nambu was to show that Hamiltonian mechanics is not the only
formulation that makes a statistical mechanics possible.
\\
A manifold $\mathbb{M}$ is called a Nambu-Poisson manifold \cite{ta} if there exits a ${\Bbb R}$-multi-linear map
\be
\{ ~,\ldots,~ \}~:~[C^{\infty }(\mathbb{M})]^{\otimes n} \rightarrow
C^{\infty }(\mathbb{M}).
\en
This is called Nambu-Poisson bracket of order $n$
 $\forall f_1 , f_2 , \ldots , f_{2n-1} \in C^{\infty }(\mathbb{M})$.
This bracket satisfies
\begin{enumerate}
\item
$\{ f_1, \ldots ,f_n \}=(-1)^{\epsilon(\sigma)}\{ f_{\sigma(1)}, \ldots ,
f_{\sigma(n)} \}$,
\item $\{ f_1 f_2, f_3, \ldots ,f_{n+1} \}=
f_1 \{f_2, f_3, \ldots , f_{n+1} \} +
\{ f_1, f_3, \ldots, f_{n+1} \} f_2$,
\item It satisfies fundamental identity, \begin{eqnarray}
\{ \{ f_1, \ldots , f_{n-1}, f_n \}, f_{n+1}, \ldots, f_{2n-1} \} +
\{ f_n, \{ f_1, \ldots, f_{n-1}, f_{n+1} \}, f_{n+2}, \ldots , f_{2n-1} \} \nonumber
\\
 +  \cdots + \{ f_n, \ldots ,f_{2n-2}, \{ f_1, \ldots , f_{n-1}, f_{2n-1} \}\}
 =  \{ f_1, \ldots , f_{n-1}, \{ f_n, \ldots , f_{2n-1} \}\}, \nonumber
\end{eqnarray}
\end{enumerate}
where $\sigma \in S_n$---the symmetric group of $n$ elements---and $\epsilon(\sigma)$ is its parity. Let $\mathbb{M}$ denote a smooth $n$-dimensional manifold and $C^{\infty }(\mathbb{M})$ the algebra of infinitely differentiable real valued functions on $\mathbb{M}$.
\\
It is known that the Nambu dynamics on a Nambu-Poisson phase space involves $n-1$ so-called Nambu-Hamiltonians $H_1, \ldots, H_{n-1} \in C^{\infty }(\mathbb{M})$ and is governed by the following equations of motion
\be
\frac {df}{dt} =  \{ f , H_1 ,\ldots,H_{n-1} \},~\forall f \in
C^{\infty }(\mathbb{M}).
\en

A solution to the Nambu-Hamilton equations of motion produces an evolution
operator $U_t$ which by virtue of the fundamental identity preserves
the Nambu bracket structure on $C^{\infty }(M)$.
A function $ f \in C^{\infty}(M)$ is a first integral of $X_{H_1, \cdots H_{n-1}}$
if and only if
$$ \{ f,H_1,H_2, \cdots ,H_{n-1} \} ~=~ 0.$$

\bigskip

We focus on {\em trilinear} antisymmetric Nambu-Poisson bracket $\{f, H_1, H_2 \}$, where 
$H_1$ and $H_2$ are a pair of Hamiltonians. Since $ \{f, H_1 \}$ represents the contraction of 
$\{f,H_1,H_2\}$ with $H_2$, the second Hamiltonian $H_2$ is always a Casimir of the original 
Poisson bracket, i.e., $\{f, H_2\} = 0$ for any $f$. It has been found \cite{nb,gu} 
that two and three dimensional Euler equations of incompressible fluid can be manifested 
in terms of Nambu-dynamics. In particular, Nambu mechanics is extended 
to incompressible ideal hydrodynamical flow using energy and helicity in 3D. It is equivalent to
\be
\frac{df}{dt} = \{f,h,H \}
\en
where $H = \frac{1}{2}\int \vec{u}\cdot \vec{u} \, d^3x $ is the Hamiltonian of the fluid with velocity 
$ \vec{u}$ and vorticity $\vec{\omega} = \vec{\nabla} \times \vec{u}$; 
$ h =  \frac{1}{2}\int \vec{u}\cdot \vec{\omega} \, d^3x $ is the helicity.
\\
In our case stream-helicity does not seem to participate in the construction of Nambu-dynamics 
by acting like a second Hamiltonian. Structurally the Lagrangian advection equation for 
velocity vector potential (\ref{6}) without $\eta$ term is very much similar 
to the incompressible 2D hydrodynamics equation. The only difference is that the equation (\ref{6})
is a three dimensional one. But this makes an enormous difference in studying Nambu-Poisson formulation
of equation (\ref{6}).

We overcome this problem by introducing a pair of stream functions.
Thus we define
\be
\vec{u} = \vec{\nabla} \psi_1 \times \vec{\nabla} \psi_2,
\en
where $\psi_1$ and $\psi_2$ are stream functions. One can easily see that
$\vec{\nabla} \cdot \vec{u} = 0$. This enables us to 
express the equation (\ref{6}) as
\be
\frac{\partial \vec{\xi}}{\partial t} + \{\vec{\xi}, \psi_1, \psi_2 \} = \textrm{curl}^{-1}\vec{\eta},
\label{2n}
\en
where $$ \{\vec{\xi}, \psi_1, \psi_2 \} = \frac{\partial(\vec{\xi}, \psi_1, \psi_2)}{\partial(x,y,z)}.$$

We assume that the derivatives of the two Hamiltonians $H_1$ and $H_2$ with respect to $\vec{\xi}$ are given by $\frac{\delta H_1}{\delta \vec{\xi}} = \psi_1$ and $\frac{\delta H_2}{\delta \vec{\xi}} = \psi_2$ respectively. We propose that the equation (\ref{2n}) could be written as
\be
\frac{\partial \vec{\xi}}{\partial t} + {\cal N} \left(\vec{\xi}, 
\frac{\delta H_1}{\delta \vec{\xi}}, \frac{\delta H_2}{\delta \vec{\xi}} \right) = \textrm{curl}^{-1}\vec{\eta}
\label{2n+1}
\en
where ${\cal N}$ is a trilinear antisymmetric bracket, which is defined earlier.

\section{CONCLUSION}
In the light of the fact that the stream-helicity can turn out to be a well-defined conserved quantity in certain kinds of fluid flow, the velocity vector potential, that basically is thought to be a rather unphysical quantity, gets a life of its own.
This is so because the velocity vector potential can interplay with the fluid velocity to construct the restrictedly conserved quantity --- stream-helicity--- which in itself is very much physical as has been demonstrated in this paper by highlighting very sound topological meaning for it.
Quite possibly, this conservation may have kept hidden within itself a symmetry that may be an interesting prospect to study in future.
This paper's naive attempt has been to initiate serious study of velocity vector potential's dynamics in fluids.
It has been showcased that this vector potential can regale itself with the mathematical structures galore {\it viz.}, quaternionic formulation, Clebsch parameterisation and Nambu structure.
To conclude, herein has been documented probably the first coherent, though diverse, study of the dynamics of the velocity vector potential in a 3D inviscid incompressible flow; hopefully more works will be done in this direction will follow soon.
\acknowledgements
Professor J.K. Bhattacharjee is heartily thanked for the various important constructive comments on this work.
The authors are indebted to Mr. Ayan Paul for providing them with various scholarly articles.
CSIR (India) is gratefully acknowledged for the financial support in the form of the fellowship awarded to the first author.
PG expresses grateful thanks to Professor J\"urgen Jost 
for gracious hospitality at the Max Planck Insitute for 
Mathematics in the Sciences.


\end{document}